\documentclass[a4,12pt]{article}

\usepackage{pstricks}
\usepackage{epsfig}
\usepackage{amsmath}
\usepackage{bm}
\usepackage{array}
\usepackage{subfig}
\usepackage{graphicx}
\usepackage{cite}
\usepackage{amssymb}

\def\om{\omega}

\def\ol{\overline}
\def\({\left(}
\def\){\right)}
\def\slo{\hspace{-6pt}/}
\def\slt{\hspace{-5pt}/}

\begin{document}
\setlength{\baselineskip}{18pt}
\vspace{-3cm}
\begin{flushright}
OSU-HEP-11-10\\
\vspace{1cm}
\end{flushright}

\renewcommand{\thefootnote}{\fnsymbol{footnote}}

\begin{center}
{\Large\bf Radiative Neutrino Mass Generation \\[0.1in]
through Vector--like Quarks}\\
\end{center}

\vspace{0.5cm}
\begin{center}
{\large \it {}~K.S. Babu\footnote{Email:
babu@okstate.edu} and J. Julio\footnote{julio.julio@okstate.edu} }
\vspace{0.5cm}

{\em Department of Physics \\ Oklahoma State University \\
Stillwater, OK 74078, USA }

\end{center}

\begin{abstract}
\setlength{\baselineskip}{18pt}

A new model of radiative neutrino masses generated via two--loop diagrams is proposed involving
a charge $2/3$ vector--like quark and a doublet of leptoquark scalars.  This model predicts one of the neutrinos
to be massless and admits both the normal and inverted neutrino mass hierarchies with correlated
predictions for $\ell_i \rightarrow \ell_j + \gamma$ branching ratios.
New contributions to CP violation in $B_s-\overline{B}_s$ mixing arise in the
model through leptoquark box diagrams, which can explain the anomalous dimuon events reported by the
D\O\ collaboration. These leptoquarks, with masses below 500 GeV, also provide a natural resolution to the apparent
discrepancy in the measured values of the CP violation parameters $\sin2\beta$ and $\epsilon_K$.

\end{abstract}

\newpage
\renewcommand{\thefootnote}{\arabic{footnote}}
\setcounter{footnote}{0}

\section{Introduction}

Neutrinos must have tiny masses, so that different flavors can oscillate among one another,
as observed in experiments.  An elegant and natural way to generate the tiny masses is through the dimension--five
lepton number violating operator ${\cal L} = \mathcal{O}_1/M$ where \cite{weinberg}
\begin{equation}
\label{o1}
\mathcal{O}_1={L^iL^jH^kH^l\,\epsilon_{ik}\,\epsilon_{jl}}~.
\end{equation}
Here $L$ is the lepton doublet and $H$ the Higgs doublet, with $i,j = 1,2$ being
$SU(2)_L$ indices. The suppression by an inverse power of $M$, which can be much
greater than the weak scale, explains the smallness of neutrino mass, which is given by $m_\nu \sim v^2/M$, with
$\left\langle H^0 \right\rangle \equiv v \simeq 174$ GeV being the Higgs boson vacuum expectation value (VEV).
Operator ${\cal O}_1$ is naturally realized through the seesaw mechanism wherein right--handed neutrinos, which
are singlets of the standard model (SM) gauge group with large
Majorana masses, are integrated out \cite{seesaw}.  The effective mass scale $M$ should be of order
$10^{14}$ GeV in order to generate neutrino masses of order 0.1 eV, as indicated by neutrino
oscillation experiments.  Such a large scale of $M$ would however make this mechanism difficult to
test directly in experiments such as the ones pursued at the Large Hadron Collider.

An alternative method for inducing naturally small neutrino masses is the
radiative mass generation mechanism \cite{chengli,zee80,hs,zee86,babu88}.  This scheme posits that the dimension 5
operator ${\cal O}_1$ of Eq. (\ref{o1}) is absent, or is highly suppressed, so that neutrino masses remain zero at the tree level.
Lepton number violation arises through effective operators with dimension $d > 5$, typically containing
charged fermions as well as the neutrino fields.  These operators can be converted to neutrino mass, but only
through loop diagrams, wherein all charged fermions are annihilated.  The induced neutrino masses are naturally
small, even when new particles needed to generate the $d > 5$ lepton number violating operators
have masses in the TeV range, owing to chirality and loop suppression factors.

The simplest set of operators carrying two units of lepton number appropriate for small Majorana neutrino
mass generation, in the absence of ${\cal O}_1$
of Eq. (\ref{o1}), is of dimension seven.  There are six such $d=7$ operators \cite{babu-leung}:
\begin{eqnarray}
\label{ops}
{\cal O}_2 &=& L^i L^j L^k e^c H^l \,\epsilon_{ij} \epsilon_{kl} \nonumber \\
{\cal O}_3 &=& \{L^i L^j Q^k d^c H^l \,\epsilon_{ij} \epsilon_{kl},\,\, L^i L^j Q^k d^c H^l \,\epsilon_{ik} \epsilon_{jl}\} \nonumber \\
{\cal O}_4 &=& \{ L^i L^j \bar{Q}_i \bar{u^c}H^k \,\epsilon_{jk},\,\, L^i L^j \bar{Q}_k \bar{u^c}H^k \,\epsilon_{ij} \} \nonumber \\
{\cal O}_8 &=& L^i \bar{e^c} \bar{u^c} d^c H^j \,\epsilon_{ij}\,.
\label{op}
\end{eqnarray}
Here the generation and color indices have been suppressed.  $Q,L$ denote left-handed quark and lepton doublets,
while $u^c,d^c,e^c$ denote left-handed anti-quark and anti-lepton singlets of the standard model.
A full list of $\Delta L = 2$ effective
operators through $d=11$ is given in Ref. \cite{babu-leung}.   Among the $d=7$ operators
of Eq. (\ref{ops}), ${\cal O}_2$  is perhaps the simplest,
which can be induced when the scalar spectrum of the standard model is extended to include a second Higgs boson doublet
and a charged singlet scalar field $h^\pm$.  This is the well--studied Zee model of neutrino masses \cite{zee80}.
In its simplest version, with natural flavor conservation in the Higgs sector, this model predicts vanishing diagonal
elements of the neutrino mass matrix \cite{zee80,wolf}, which is now excluded by neutrino oscillation data \cite{zee-exp}.

A second widely studied model of radiative neutrino masse generation \cite{zee86,babu88} has a purely leptonic effective $d=9$
operator, ${\cal O}_9 = L^i L^j L^k e^c L^l e^c \,\epsilon_{ij} \epsilon_{kl}$, suppressed by $M^{-5}$.  Here neutrino masses
are induced via two--loop diagrams.  This operator can be
obtained when the standard model is extended to include a singly charged ($h^+$) scalar and a doubly charged ($k^{++}$) scalar.
The resulting model fits the neutrino oscillation data well, and also predicts a host of leptonic flavor violation processes, some
of which within reach of ongoing and next generation experiments \cite{zee-babu-phen}.  Operator ${\cal O}_8$ of Eq. (\ref{op})
is best induced by scalar leptoquarks, as recently shown by us in Ref. \cite{bj}.  This model leads to consistent neutrino phenomenology and interesting flavor effects in both the quark and the lepton sectors \cite{bj}.  For discussions of models based on other operators, see
Ref. \cite{babu-leung,choi,degouvea}.

Operator ${\cal O}_3$ of Eq. (\ref{op}) is the main focus of this paper.  It has two different $SU(2)_L$ contractions possible, as shown in Eq. (\ref{op}).  These operators arise in supersymmetric models with $R$--parity violation.  The superpotential couplings $W' \supset \lambda L L e^c + \lambda' Q L d^c$
would generate ${\cal O}_3$ once the SUSY particles are integrated out \cite{hs,dey}.  The $QLd^c$ term would induce the second contraction
of ${\cal O}_3$ in Eq. (\ref{op}), the $LLe^c$ term would induce ${\cal O}_2$, while the product of $QLd^c$ and $LLe^c$ would induce the first contraction of ${\cal O}_3$.  There is an important difference in the first and second $SU(2)_L$ contractions of ${\cal O}_3$:  In
the second contraction, neutrino masses are induced at the one loop, while in the first contraction, they arise only at the two loop
level.  (In the second contraction, two neutrino fields appear, while the first has one neutrino field and a charged
lepton field, which must be annihilated to convert this operator to neutrino mass.)  The focus of this paper is models which induce the first contraction of ${\cal O}_3$, without inducing other operators
that lead to one loop neutrino masses.  SUSY with $R$--parity violation does not fit this requirement, as ${\cal O}_2$ and/or
the second contraction of ${\cal O}_3$ are also induced there.  The simplest possibility we have found is to add a vector--like
charge $2/3$ iso--singlet quark to the SM, along with a doublet of leptoquark scalars.  The induced neutrino mass is of the form
\begin{equation}
m_\nu \sim \frac{f g  h \lambda_b}{(16\pi^2)^2}\,\,\frac{v^2}{M}~,
\label{estimate}
\end{equation}
where $f,g,h$ are dimensionless Yukawa couplings, $\lambda_b = m_b/v$ is the $b$--quark Yukawa coupling, and $M$ stands for
an effective mass of the vector--like quark/leptoquark.  For $f \sim g \sim h \sim 10^{-2}$, the mass scale $M$ should be
of order TeV, in order to generate $m_\nu \sim 0.1$ eV.  It is, however, evident from Eq. (\ref{estimate}) that $M$ can be
as large as about $10^8$ GeV, when $f,g,h$ are of order one.  There are several reasons for considering low values of
$M$, first and foremost being direct tests of the vector quark and the leptoquarks at the LHC.  There are hints of new physics in the $B$ meson system, which can be explained by the new leptoquark scalars and/or the vector--like quark of the present model.  The D\O\ collaboration
has reported an excess in the same sign di-muon asymmetry in $B$ decays \cite{abazov}, which may be a hint for new CP violation
in $B_s-\overline{B}_s$ mixing.  There has also been a tension in the determinations of the CP asymmetry parameters
$\sin2\beta$ in $B$ meson decay and $\epsilon_K$ in Kaon decay, which may need new physics \cite{lunghi-soni}.
The present model, with leptoquark masses below a TeV, can explain these anomalies.  Furthermore, when this model
is eventually embedded in a supersymmetric framework, $M$ of Eq. (\ref{estimate}) will have to be close to the SUSY
breaking scale, owing to the SUSY non-renormalization theorem, with the consequence that all loop diagrams that generate neutrino
masses cancel in the supersymmetric limit.

This rest of the paper is organized as follows. In Sec. 2 we present the model leading
to the two-loop neutrino mass generation via ${\cal O}_3$, the first contraction of Eq. (\ref{op}).
In Sec. 3 we obtain the experimental constraints on the model parameters arising from rare process
in the quark as well as lepton sectors. Here we show how the proposed model explains
the discrepancy observed by D\O\ in the CP asymmetry of the $B_{s}$ system. New contributions
to the CP violating decay $B_d \rightarrow J/\Psi K_S$ are shown to be of the right magnitude to explain the
apparent tension between $\sin2\beta$ and $\epsilon_K$ determination. In Sec. 3  we also evaluate
the rate for neutrinoless double beta
decay induced via the vector-scalar exchange mechanism \cite{babu-mohapatra}.
Finally, we give our conclusions in Sec. 4.

\section{Radiative neutrino mass model with vector--like quark}

We wish to generate the operator $(L \cdot L)(Q \cdot H)d^c$ in a renormalizable theory.  Here and in discussions that follow
we use a compact dot product notation for $SU(2)_L$ contraction: $L \cdot L = L^i L^j \epsilon_{ij}$, $Q \cdot H = Q^i H^j \epsilon_{ij}$, etc.
The simplest way to generate this operator, without inducing other operators that generate neutrino masses at one loop, is by
integrating out a charge $2/3$ iso-singlet vector--like quark, and a doublet of scalar leptoquarks.
These fields transform under $SU(3)_c \times SU(2)_L \times U(1)_Y$ as
\begin{equation}
{\rm Fermions:} \quad U(3,1,2/3) +  U^c(3^*,1,-2/3),~~~~ {\rm Scalars:}~ \Omega(3,2,1/6) \equiv \left(\begin{array} {c} \omega^{2/3} \\ \omega^{-1/3} \end{array}
\right)~.
\label{new_fields}
\end{equation}
These particles will have new Yukawa interactions with the SM fermions as well as gauge invariant masses given by
\begin{equation}
\mathcal{L}_Y^{\rm new} = \( g_{ij} d^c_j L_i\cdot\Omega ~+~ h_i U L_i \cdot
\tilde{\Omega} ~-~ f_i U^c Q_i \cdot H ~+~ {\rm h.c.} \) - M U U^c,
\label{new_int}
\end{equation}
where $\tilde{\Omega} \equiv i \tau^2 \Omega^*$. Here the dots indicates $SU(2)_L$
contraction, as mentioned earlier, and we use indices $i,j$ to denote generations.
Possible mass terms $m_i\,u_i^c\, U$, not shown in Eq. (\ref{new_int}), can be rotated away by field redefinitions.
The simultaneous presence of the interaction terms $g_{ij}, h_i, f_i$  would lead to lepton number
violation by two units, a necessary condition for neutrino mass generation.

We should also specify the scalar interactions that couple the leptoquark $\Omega$ with the SM Higgs doublet $H$.
There is a single non-trivial quartic coupling between these two fields:
\begin{eqnarray}
\mathcal{L}_{\rm quart}^{\rm new} &=& \lambda \left|\Omega \cdot H \right|^2
\label{new_quart}
\end{eqnarray}
When the neutral component of the SM Higgs doublet $H^0$ acquires a VEV, this quartic
coupling will generate a mass splitting between $\om^{2/3}$ and $\om^{-1/3}$ leptoquarks:
\begin{equation}
M_{\om^{-1/3}}^2 \equiv M_1^2, ~~~
M_{\om^{2/3}}^2 \equiv M_2^2 = M_1^2 - \lambda v^2,
\end{equation}
where $v\equiv \sqrt{2}m_W/g \simeq 174$ GeV.

The mass matrix for the charge $2/3$ quarks, including $U,U^c$ fields, that follows from Eq. (\ref{new_int})
has the form
\begin{equation}
M_u = \left( \begin{array}{cc}
Y_uv & 0 \\ fv & M \end{array} \right),
\end{equation}
where $(u_i^c, U^c)$ multiply on the left and $(u_i, U)$ multiply on the right.
Here $Y_u$ is a $3\times 3$ Yukawa coupling matrix,  $f$ is a $1\times
3$ row vector, and $0$ stands for the $3 \times 1$ null column vector. This mass matrix can be
diagonalized by a biunitary transformation
\begin{equation}
M^{\rm d}_u = U M_u V^\dagger
\end{equation}
where $U,V$ are $4 \times 4$ unitary matrices.
Without loss of generality we choose a basis where the $3 \times 3$ matrices for the down
quarks and charged leptons are diagonal.
Thus, the CKM matrix will be the $4\times 3$ sub-matrix of the $4 \times 4$ matrix $V$.
The charged current interactions of the quarks, therefore, become
\begin{eqnarray}
\mathcal{L}^{cc,q}_{\rm vector} &=& \frac{g}{2\sqrt{2}}~\ol{u}_\alpha V_{\alpha i}
\gamma^\mu(1-\gamma_5) d_i W_\mu^+ + {\rm h.c.},
\label{cc-vec}
\\
\mathcal{L}^{cc,q}_{\rm scalar} &= & \frac{g}{2\sqrt{2}m_W}~\ol{u}_\alpha \left[
(M^d_u)_{\alpha}V_{\alpha i}
(1-\gamma_5) - V_{\alpha i}(M_d)_{i} (1+\gamma_5)
\right] d_i \,H^+ ~+~{\rm h.c.}
\label{cc-sc}
\end{eqnarray}
The Greek indices $\alpha,\beta=1-4$ label generations in the up--quark sector
($u_1=u,~u_2=c,~u_3=t,~u_4=t'$),
while the Latin indices $i,j=1-3$ label generations in the down--quark and lepton
sectors.
Introduction of vector-like quarks $U,U^c$ to the SM spectrum will induce flavor
changing neutral currents (FCNC) in the charge $2/3$ quark sector, which are given by
\begin{eqnarray}
\mathcal{L}^{nc,q}_{\rm vector} &=& \frac{g}{4\cos{\theta_{W}}}
~\ol{u}_{\alpha} \left[
\delta_{\alpha\beta}\gamma^\mu\left(1-\tfrac{8}{3}\sin^2\theta_W - \gamma_5 \right) -
V_{\alpha 4}V^*_{\beta 4}
\gamma^\mu (1-\gamma_5)
\right]
u_\beta Z_\mu ,
\label{nc-vec}
\\
\mathcal{L}^{nc,q}_{\rm scalar} &=&
\frac{g}{2\sqrt{2}m_W} ~\ol{u}_\alpha (M_u)_\alpha V_{\alpha j}
V^*_{\beta j} (1-\gamma_5) u_{\beta} H^0 + {\rm h.c.}
\label{nc-sc}
\end{eqnarray}
These interactions can generate tree-level $D-\ol{D}$  mixing, as discussed in the next section,
which will strongly constrain the product $|V_{14}V_{24}|$.

The $4 \times 4$ unitary matrix $V$
can be parameterized as \cite{mix-par}
\vspace*{0.05in}

\begin{footnotesize}
\begin{eqnarray}
V &=& \nonumber \\
&& \hspace{-1cm}\left(
\begin{array}{cccc}
c_{12} c_{13} c_{14} & c_{13} c_{14} s_{12} &  c_{14} s_{13}e^{-i \delta _{13}} &
s_{14} e^{-i \delta _{14}} \\ \\
-c_{23} c_{24} s_{12}-c_{12} c_{24} s_{13} s_{23} e^{i \delta _{13}} &
c_{12} c_{23} c_{24}-c_{24} s_{12} s_{13} s_{23} e^{i \delta _{13}} &
c_{13} c_{24} s_{23} &
c_{14} s_{24} e^{-i \delta _{24}} \\
-c_{12} c_{13} s_{14} s_{24} e^{i(\delta _{14}-\delta _{24})} &
-c_{13} s_{12} s_{14} s_{24} e^{i (\delta _{14}- \delta _{24})} &
-s_{13} s_{14} s_{24} e^{-i(\delta _{13}+\delta _{24}- \delta _{14})} & \\ \\
c_{34} s_{12} s_{23}-c_{12} c_{23} c_{34} s_{13} e^{i \delta _{13}} &
-c_{12} c_{34} s_{23}-c_{23} c_{34} s_{12} s_{13} e^{i \delta _{13}} &
c_{13} c_{23} c_{34} &
c_{14} c_{24} s_{34} \\
-c_{12} c_{13} c_{24} s_{14} s_{34} e^{i \delta _{14}} &
-c_{12} c_{23} s_{24} s_{34} e^{i \delta _{24}} &
-c_{13} s_{23} s_{24} s_{34} e^{i \delta _{24}} &  \\
+c_{23} s_{12} s_{24} s_{34} e^{i \delta _{24}} &
-c_{13} c_{24} s_{12} s_{14} s_{34} e^{i \delta _{14}} &
-c_{24} s_{13} s_{14} s_{34} e^{i( \delta _{14}- \delta _{13})} &  \\
+c_{12} s_{23} s_{24} s_{34} s_{13} e^{i (\delta _{13}+ \delta _{24})} &
+s_{12} s_{23} s_{24} s_{34} s_{13} e^{i (\delta _{13}+ \delta _{24})} & & \\ \\
-c_{12} c_{13} c_{24} c_{34} s_{14} e^{i \delta _{14}} &
-c_{12} c_{23} c_{34} s_{24} e^{i \delta _{24}}+c_{12} s_{23} s_{34} &
-c_{13} c_{23} s_{34} &
c_{14} c_{24} c_{34} \\
+c_{12} c_{23} s_{13} s_{34} e^{i \delta _{13}} &
-c_{13} c_{24} c_{34} s_{12} s_{14} e^{i \delta _{14}} &
-c_{13} c_{34} s_{23} s_{24} e^{i \delta _{24}} & \\
+c_{23} c_{34} s_{12} s_{24} e^{i \delta _{24}}-s_{12} s_{23} s_{34} &
+c_{23} s_{12} s_{13} s_{34} e^{i \delta _{13}} &
-c_{24} c_{34} s_{13} s_{14} e^{i (\delta _{14}- \delta _{13})} & \\
+c_{12} c_{34} s_{13} s_{23} s_{24} e^{i (\delta _{13}+ \delta _{24})} &
+c_{34} s_{12} s_{13} s_{23} s_{24} e^{i (\delta _{13}+ \delta _{24})} & &
\end{array}
\right)
,\nonumber \\
&&
\label{u-par}
\end{eqnarray}
\end{footnotesize}

\vspace*{-0.1in}
\noindent where $s_{\alpha\beta}\equiv \sin\theta_{\alpha\beta},
~c_{\alpha\beta}\equiv\cos\theta_{\alpha\beta}$. The CKM mixing matrix elements $V_{\alpha i}$ are
the elements of the $4\times 3$ sub-matrix of $V$.  In terms of the fermion mass
eigenstates,  Eq. (\ref{new_int}) can be written as
\begin{eqnarray}
\mathcal{L}_Y^{\rm new} &=& \bar{d}_j ~(g^T)_{ji}~\frac{(1-\gamma_5)}{2}~
\left(\nu_{i}\omega^{-1/3} - \ell_i \omega^{2/3} \right) \nonumber \\
&& - ~\left( \nu^T_i C^T \omega^{-2/3} + \ell^T_i C^T \omega^{1/3} \right)
h_i~V^*_{\alpha 4} \frac{(1-\gamma_5)}{2} u_\alpha + {\rm h.c.}
\label{new_int1}
\end{eqnarray}
which will be used in our calculations.

\subsection{Two-loop neutrino masses}

\begin{figure}
\centering
	\includegraphics[scale=0.85]{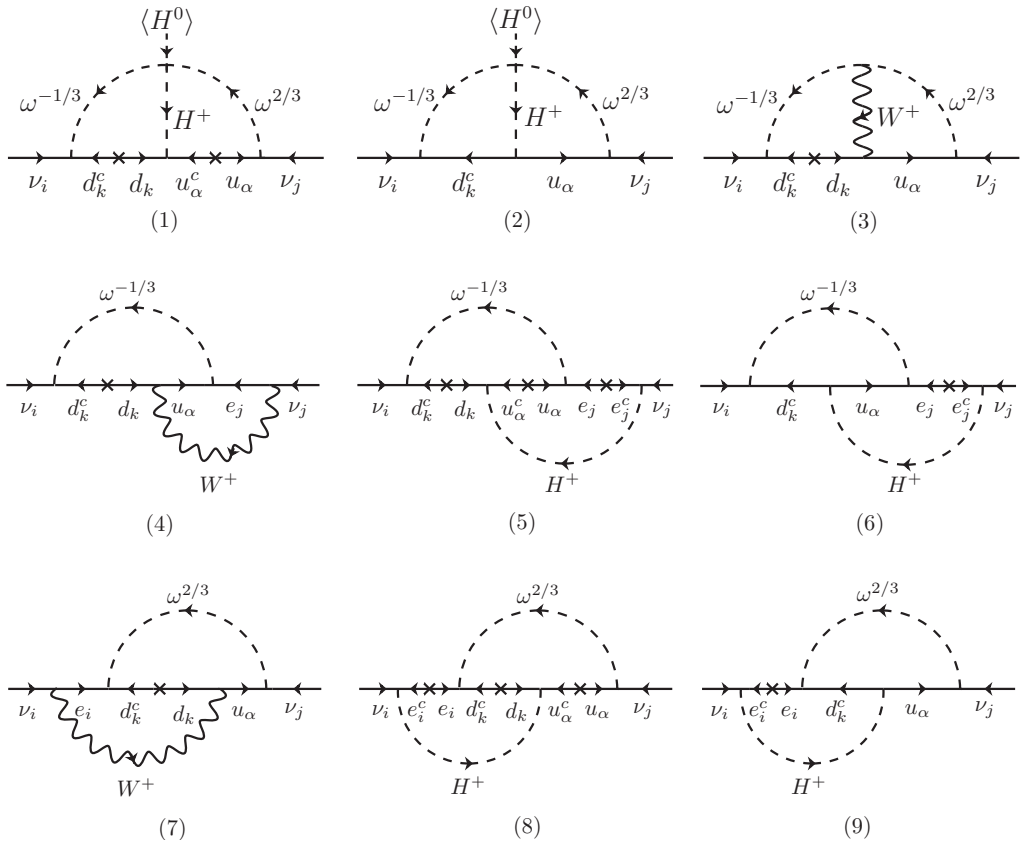}
	\caption{Two-loop diagrams leading to finite neutrino masses in general $R_\xi$ gauge.}
\label{two-loop}
\end{figure}

By combining the interactions given in Eqs. (\ref{new_quart}), (\ref{cc-vec}), (\ref{cc-sc}) and (\ref{new_int1}),
one can construct diagrams generating neutrino masses.  These diagrams arise at the two loop level, and are
shown in Fig. \ref{two-loop}.  We have done the evaluation of these diagrams in general $R_\xi$ gauge, so the
unphysical Goldstone mode $H^+$ also appear in this set.  A non-trivial check of the calculation is
the gauge independence of the induced neutrino mass, which we will show explicitly.
Since the external neutrinos are Majorana particles, there is another set of diagrams identical to the ones
in Fig. \ref{two-loop}, but with all internal particles replaced by their charge conjugates.  The sum of these
diagrams would make the neutrino mass matrix symmetric in flavor space.

The induced neutrino mass matrix is proportional to the down quark mass matrix, since these diagrams make use of the SM charged currents,
which require a chirality flip for the $d^c$ fields.  This is explicitly shown in Fig. \ref{two-loop}.
The neutrino mass matrix, therefore, can be written as
\begin{eqnarray}
(M_\nu)_{ij} = \frac{3}{2}g^2 m_b  \left[h_i (V^\dagger)_{4\alpha}V_{\alpha
    k}(D_d)_k(g^T)_{k j} \hat{I}_{\alpha kij}
    + g_{ik}(D_d)_{k}(V^T)_{k\alpha}(V^*)_{\alpha 4}h_j \hat{I}_{\alpha kji}\right],
\label{nu-mass1}
\end{eqnarray}
where 3 is a color factor and $D_d$ is the normalized down quark mass matrix,
\begin{eqnarray}
D_d = {\rm diag}\left[\frac{m_d}{m_b},\frac{m_s}{m_b},1 \right]~.
\end{eqnarray}
The function $\hat{I}_{\alpha kij}$ is a sum of loop integrals defined as
\begin{equation}
\hat{I}_{\alpha kij} = \sum_{n=1}^{3}\hat{I}_{\alpha k}^{(n)}
+ \sum_{n=4}^6\hat{I}_{\alpha ki}^{(n)} + \sum_{n=7}^9\hat{I}_{\alpha kj}^{(n)}\,,
\end{equation}
where the integral $\hat{I}^{(n)}$ is given by\footnote{Owing to the unitarity of
$V$, only terms containing $m_{u_\alpha}$ are relevant in
generating neutrino mass (see Eq.
(\ref{nu-mass1})). All
terms that are independent of $m_{u_\alpha}$ will add up to zero, and therefore, such terms are
not written explicitly.}
\begin{footnotesize}
\begin{eqnarray}
\hat{I}^{(1)}_{\alpha k} + \hat{I}^{(2)}_{\alpha k} &=&
\(\frac{M_2^2-M_1^2}{m_W^2}\)~
\int \frac{d^4k}{(2\pi)^4}~\frac{d^4q}{(2\pi)^4}
~\frac{(k\slo+q\slt)k\slo}{k^2}
\( 1 + \xi \frac{m_W^2}{k^2-\xi m_W^2} \) \nonumber \\ \nonumber \\
&& \times ~\frac{1}{q^2-M_1^2}~\frac{1}{q^2-m_{d_k}^2}~\frac{1}{(k+q)^2-M_2^2}
~\frac{1}{(k+q)^2-m_{u_\alpha}^2},
\label{int1}
\\ \nonumber \\ \nonumber \\
\hat{I}^{(3)}_{\alpha k} &=&
\int \frac{d^4k}{(2\pi)^4}~\frac{d^4q}{(2\pi)^4}
~\frac{k\slo+q\slt}{k^2-m_W^2}
\left[ -k\slo - 2q\slt + \frac{k\slo~k\cdot(k+2q)}{k^2}
\( 1 - \xi \frac{k^2-m_W^2}{k^2-\xi m_W^2} \) \right]\nonumber \\
\nonumber \\
&& \times ~\frac{1}{q^2-M_1^2}~\frac{1}{q^2-m_{d_k}^2}~\frac{1}{(k+q)^2-M_2^2}
~\frac{1}{(k+q)^2-m_{u_\alpha}^2},
\label{int2}
\\ \nonumber \\ \nonumber \\
\hat{I}^{(4)}_{\alpha ki} &=&
\int \frac{d^4k}{(2\pi)^4}~\frac{d^4q}{(2\pi)^4}
~\left[ 4k\slo(k\slo+q\slt) - (k\slo+q\slt)k\slo
\(1 - \xi \frac{k^2-m_W^2}{k^2-\xi m_W^2} \)
\right]
\frac{1}{k^2-m_W^2}~\frac{1}{k^2-m_{e_i}^2}\nonumber \\
\nonumber \\
&& \times ~\frac{1}{q^2-M_1^2}~\frac{1}{q^2-m_{d_k}^2}
~\frac{1}{(k+q)^2-m_{u_\alpha}^2},
\label{int3}
\end{eqnarray}
\begin{eqnarray}
\hat{I}^{(5)}_{\alpha ki} + \hat{I}^{(6)}_{\alpha ki} &=&
-\(\frac{m_{e_i}}{m_W}\)^2 \int \frac{d^4k}{(2\pi)^4}~\frac{d^4q}{(2\pi)^4}
~\frac{(k\slo+q\slt)k\slo}{k^2}
\( 1 + \xi \frac{m_W^2}{k^2-\xi m_W^2} \) \frac{1}{k^2-m_{e_i}^2} \nonumber \\
\nonumber \\
&& \times ~\frac{1}{q^2-M_1^2}~\frac{1}{q^2-m_{d_k}^2}~\frac{1}{(k+q)^2-m_{u_\alpha}^2},
\label{int4}
\\ \nonumber \\ \nonumber \\
\hat{I}^{(7)}_{\alpha kj} &=&
\int \frac{d^4k}{(2\pi)^4}~\frac{d^4q}{(2\pi)^4}
~\frac{(k\slo+q\slt)k\slo}{k^2-m_W^2}
\( 3 - \xi \frac{k^2-m_W^2}{k^2-\xi m_W^2} \)\frac{1}{k^2-m_{e_j}^2} \nonumber \\
\nonumber \\
&& \times ~\frac{1}{q^2-m_{d_k}^2}~\frac{1}{(k+q)^2-m_{u_\alpha}^2}
~\frac{1}{(k+q)^2-M_2^2},
\label{int5}
\\ \nonumber \\ \nonumber \\
\hat{I}^{(8)}_{\alpha kj} + \hat{I}^{(9)}_{\alpha kj} &=&
\(\frac{m_{e_j}}{m_W}\)^2 \int \frac{d^4k}{(2\pi)^4}~\frac{d^4q}{(2\pi)^4}
~\frac{(k\slo+q\slt)k\slo}{k^2}
\( 1 + \xi \frac{m_W^2}{k^2-\xi m_W^2} \) \frac{1}{k^2-m_{e_j}^2} \nonumber \\
\nonumber \\
&& \times ~\frac{1}{q^2-m_{d_k}^2}~\frac{1}{(k+q)^2-m_{u_\alpha}^2}
~\frac{1}{(k+q)^2-M_2^2}.
\label{int6}
\end{eqnarray}
\end{footnotesize}

It is straightforward to show that all terms containing the
gauge parameter $\xi$ in Eqs. (\ref{int1})-({\ref{int6})
add up to zero. This means that the neutrino mass matrix elements, which are all
physical, are gauge independent.
An interplay of all diagrams of Fig. \ref{two-loop} is required to see
this gauge independence, although this can be inferred before doing
the momentum integrals.  Note that the contributions to these integrals proportional to charged
lepton masses are strongly suppressed as can be seen from Eqs. (\ref{int3})-(\ref{int6}).
Thus, it is a good approximation to work in the limit $m_{e_i} \simeq 0$. In this limit,
the neutrino mass matrix is reduced to a rank two matrix with a suppressed determinant
${\rm det}~(M_\nu) \ll \(0.01~{\rm eV}\)^3$. Thus, we have a prediction that
the lightest neutrino is essentially massless. For the purpose of evaluating these integrals we can
also set the down quark masses to zero.
Thus, the neutrino mass matrix of Eq. (\ref{nu-mass1}) can be written as
\begin{eqnarray}
\(M_\nu\)_{ij} \simeq \frac{3}{2}g^2m_b\left[h_i (V^\dagger)_{4\alpha}V_{\alpha k}(D_d)_k(g^T)_{k j}
 + g_{ik}(D_d)_k V_{k\alpha}V^*_{\alpha 4} h_j\right]I_{\alpha},
\end{eqnarray}
with $I_\alpha \equiv \hat{I}_{\alpha kij}(m_{d_k}\simeq m_{e_i}\simeq m_{e_j}\simeq0)$.
The asymptotic behavior of
integral $I_{\alpha}$ (in the limit $M_1=M_2$) is given by
\begin{eqnarray}
I_{\alpha } = \left\{ \begin{array}{l}
\(\tfrac{m_{u_\alpha}}{M_1}\)^2
\left[ -6~{\rm ln}\( \frac{m_{u_\alpha}}{M_1} \)^2 - \frac{\pi^2}{2}
+\frac{9}{2} \right]
-\pi^2 + \tfrac{15}{2} ,~{\rm for}~M_1 \gg m_{u_\alpha},m_{W}, \\ \\
\tfrac{3}{2}~{\rm ln}^2 \( \tfrac{m_{u_\alpha}}{M_1}\)^2
+ \frac{\pi^2}{2} + 6, ~{\rm for}~
m_{u_\alpha} \gg M_1, \\ \\
\(6-\frac{\pi^2}{4}\)\left[\(\frac{m_{u_\alpha}}{M_1}\)^2-1\right]
-\tfrac{\pi^2}{2}+9,~{\rm
for}~m_{u_\alpha} \sim M_1 \gg m_W.
\end{array}
\right.
\end{eqnarray}
These expressions are very helpful, especially for analytic approximations of
the integrals where the internal quarks are light quarks, and also as cross checks
of the exact numerical calculations.

The neutrino mass matrix can now be written down:
\begin{eqnarray}
M_\nu = m_0\( \begin{array}{ccc}
x & \tfrac{1}{2}\tfrac{h_1}{h_2}y + \tfrac{1}{2}\tfrac{h_2}{h_1}x &
\tfrac{1}{2}\tfrac{h_1}{h_3} + \tfrac{1}{2}\tfrac{h_3}{h_1} \\ \\
\tfrac{1}{2}\tfrac{h_1}{h_2}y + \tfrac{1}{2}\tfrac{h_2}{h_1}x & y &
\tfrac{1}{2}\tfrac{h_2}{h_3} + \tfrac{1}{2}\tfrac{h_3}{h_2}y \\ \\
\tfrac{1}{2}\tfrac{h_1}{h_3} + \tfrac{1}{2}\tfrac{h_3}{h_1}x &
\tfrac{1}{2}\tfrac{h_2}{h_3} + \frac{1}{2}\tfrac{h_3}{h_2}y & 1
\end{array}
\),
\label{nu-mass2}
\end{eqnarray}
where
\begin{eqnarray}
m_0 &\equiv& \frac{3g^2 m_b}{(16\pi^2)^2} h_3 F_3; \quad
x \equiv \(\frac{h_1F_1}{h_3F_3}\); \quad
y \equiv \(\frac{h_2F_2}{h_3F_3}\); \nonumber \\
F_j &\equiv& g_{jk}(V^\dagger)_{4\alpha} V_{\alpha k} (D_d)_k I_{\alpha}~,
\label{numass}
\end{eqnarray}
with repeated indices assumed to be summed.
By using the unitarity of the mixing matrix $V$, and the fact that $m_u,m_c \ll m_t, m_{t'}$
we have
\begin{eqnarray}
V^*_{\alpha 4}V_{\alpha k}I_{\alpha} \simeq V^*_{34}
V_{3k} \(I_{3}-I_{1}\) + V^*_{44}
V_{4k} \(I_{4}-I_{1}\).
\end{eqnarray}
Plots of $I_{3}-I_{1}$ and
$I_{4}-I_{1}$ as function of the leptoquark mass $M_1$ are shown in Fig. \ref{int-plot}, for a fixed value of the vector-quark
mass of 600 GeV.

To illustrate the range of parameters allowed by the neutrino mass, let us assume $g_{jk} \ll g_{j3},~k=1,2$, so that only $g_{j3}$ contribute to the neutrino mass matrix. If we further assume that only the the top quark (among $u,c,t$) mixes significantly with the vector-like quark, i.e.,
$f_1,f_2 \ll f_3$, then $V^*_{34} V_{33} \simeq -V^*_{44} V_{43} \simeq f_3 v/M$. Therefore, we can write
\begin{eqnarray}
F_j \simeq \frac{g_{j3} f_3 v}{M} \(I_{4}-I_{3}\).
\end{eqnarray}
For normal neutrino mass hierarchy, $m_0 \simeq 0.03~{\rm eV}$ is needed, which in turn requires
$h_3F_3 \simeq 10^{-7}$. This means that for order one values of the Yukawa
couplings $h_i, f_3, g_{j3}$, the mass of the vector-like quark and/or the leptoquarks is of
order $10^8$ GeV. Conversely, if both the LQ and vector-like quark have masses of order TeV, and if $h_3 \sim 1$, one must have $g_{33}f_3
\sim 10^{-5}$. In both regimes, lepton flavor violation processes do not strongly constrain the model parameters.  Interesting new
effects will arise, however, if the vector-like quark/leptoquark masses are near a TeV, and if some of the Yukawa couplings lie
in the range $10^{-2}-1$, as will be discussed in the next section.

\begin{figure}[t]
\centering
	\includegraphics[scale=1.]{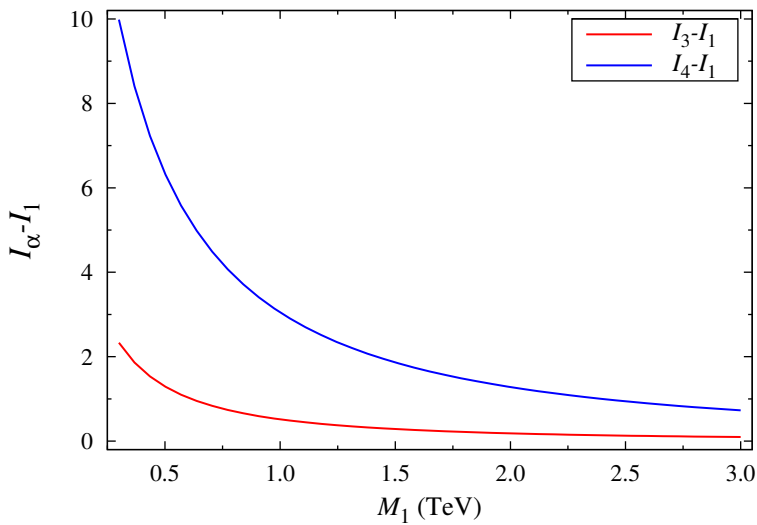}
	\caption{ Plots of the integral functions $I_{3}-I_{1}$ and $I_{4}-I_{1}$ versus the leptoquark mass $M_1$.
    The mass of the vector-like quark is taken here to be 600 GeV, with $M_1-M_2=60$ GeV.}
	\label{int-plot}
\end{figure}

Although the model predicts the lightest neutrino to be essentially massless, owing to the highly suppressed
determinant of $M_\nu$, Eq. (\ref{numass}) does admit both the normal hierarchy (NH) and the inverted hierarchy (IH)
of neutrino masses.
Since the off-diagonal elements of $M_\nu$ are uniquely related to
the diagonal elements, one can determine the values of $h_i/h_j$ for $i<j$ as
\begin{eqnarray}
\frac{h_i}{h_j} &=& \frac{(M_\nu)_{ij}}{(M_\nu)_{jj}} \left[ 1 \pm
\sqrt{1 - \frac{(M_\nu)_{jj}(M_\nu)_{ii}}{(M_\nu)_{ij}^2}}~ \right],
\label{h-ratio}
\end{eqnarray}
where $(M_\nu)_{ij}$ are obtained from
\begin{equation}
M_\nu = U_{\rm PMNS}^*(M_\nu)_{\rm diag}U_{\rm PMNS}^\dagger.
\end{equation}
Here $U_{\rm PMNS}$ is the leptonic mixing matrix
parameterized as in Ref. \cite{pdg}, while $(M_\nu)_{\rm diag}$ is given by
\begin{eqnarray}
(M_\nu)_{\rm diag} &=& {\rm diag}~\( 0, m_2e^{i\alpha},m_3 \),~{\rm for}~
{\rm NH}, \nonumber \\
(M_\nu)_{\rm diag} &=& {\rm diag}~\( m_1, m_2e^{i\alpha},0 \),~{\rm for}~
{\rm IH}.
\end{eqnarray}
Take for example the ratio $h_1/h_3$. Its value can be determined from Eq.
(\ref{h-ratio}), but this must match the product
$(h_1/h_2) \cdot (h_2/h_3)$, also obtained from the same equation.
Now, by using the central values of the current neutrino oscillation data,
$\Delta m^2_{\rm sol}=7.59\times 10^{-5}~{\rm eV^2}$,
$\Delta m^2_{\rm atm} = 2.3\times 10^{-3}~{\rm eV^2}$, $\sin^2\theta_{12}=0.304$,
$\sin^2\theta_{23}=0.5$, and the upper limit on $\theta_{13}$,
$\sin^2\theta_{13} \leq 0.04$ \cite{schwetz}, one can find the allowed
values of $h_1/h_3$.  This result is plotted in Fig. \ref{fitting} versus $\sin^2\theta_{13}$, both for NH (upper left panel)
and for IH (upper right panel). From this
figure we see that the ratio of $h_2/h_3$ has to be of order one for normal hierarchy,
while it can range from 0.5 to 1000 for inverted hierarchy. In both cases, the value of $\theta_{13}$
is allowed to range from zero up to the current upper limit. Recently, the T2K experiment \cite{t2k} has
reported an indication of nonzero $\theta_{13}$, with the best fit value (assuming $\sin^2 2\theta_{23}=1$
and $\delta=0$) being $\sin^2 2 \theta_{13}=0.11~(0.14)$ for normal (inverted) hierarchy.  MINOS experiment
also finds supporting evidence, although with less significance \cite{MINOS}. The present model can accommodate these
indications for a sizable $\theta_{13}$.

\begin{figure}[t]
\centering
	\includegraphics[scale=0.78]{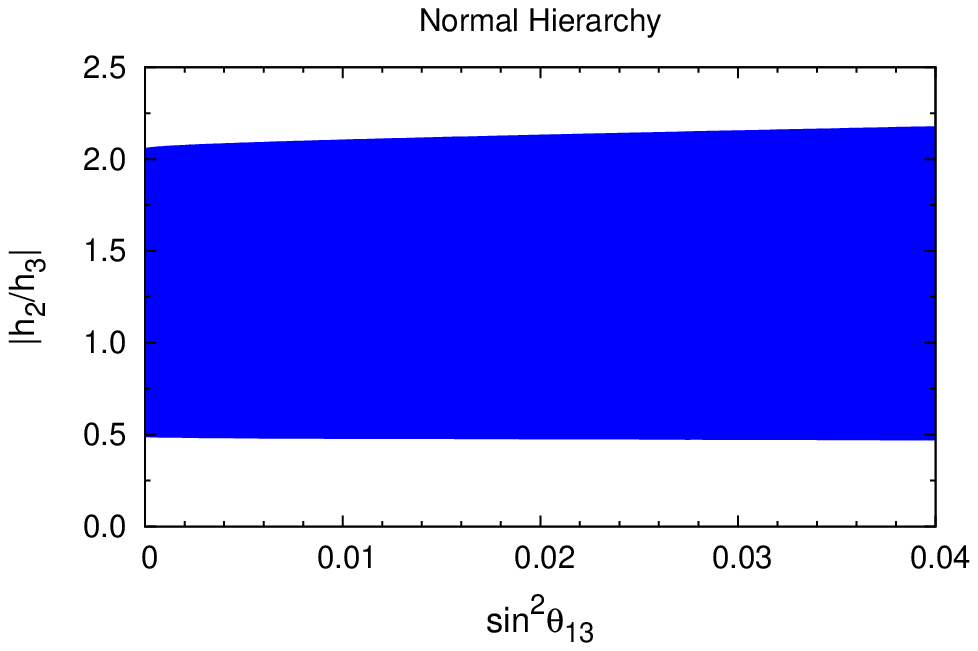}
	\includegraphics[scale=0.78]{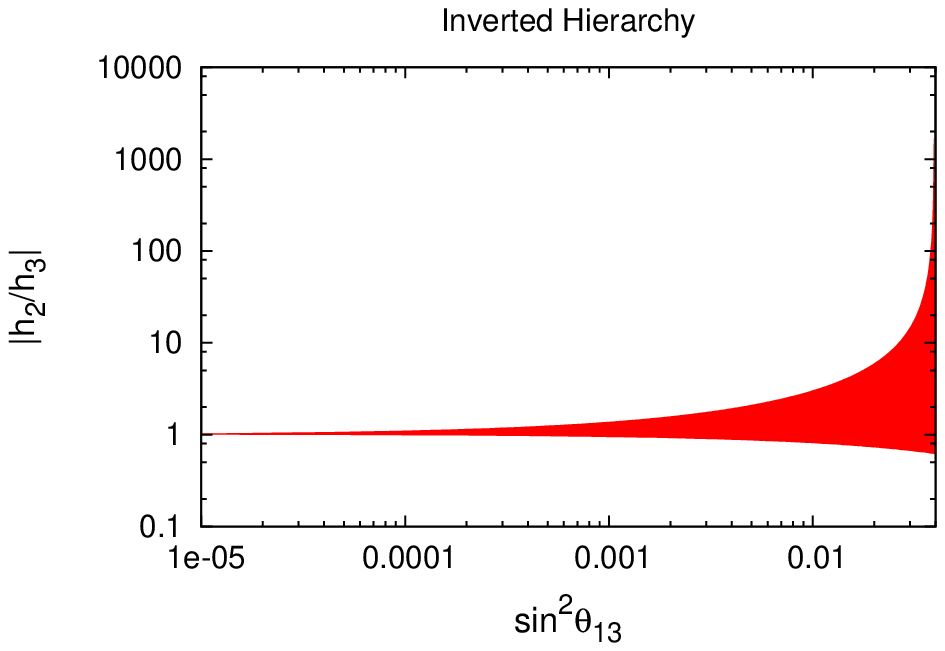}
	\includegraphics[scale=0.78]{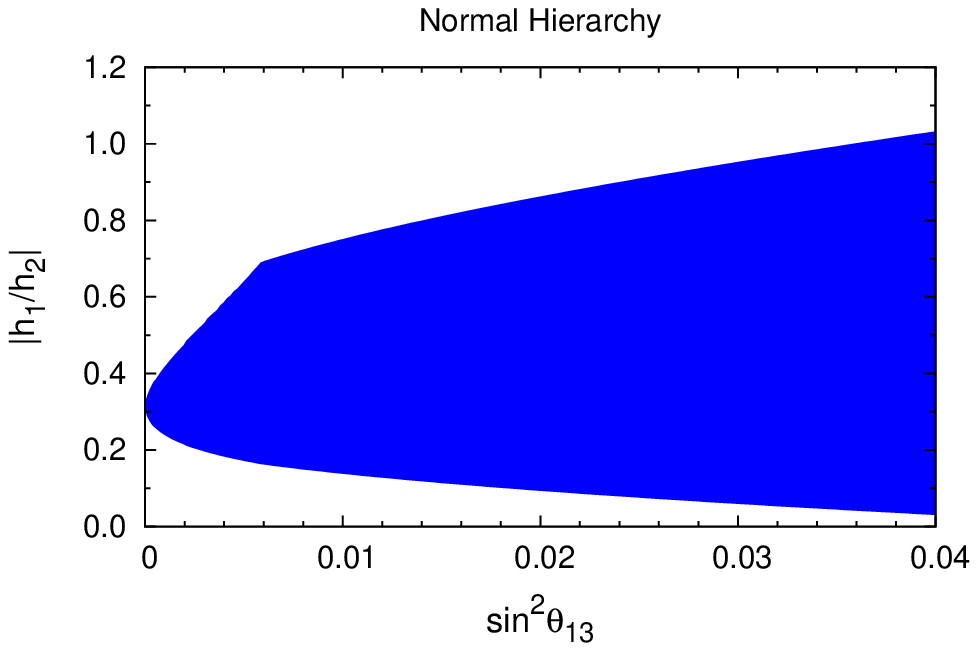}
	\includegraphics[scale=0.78]{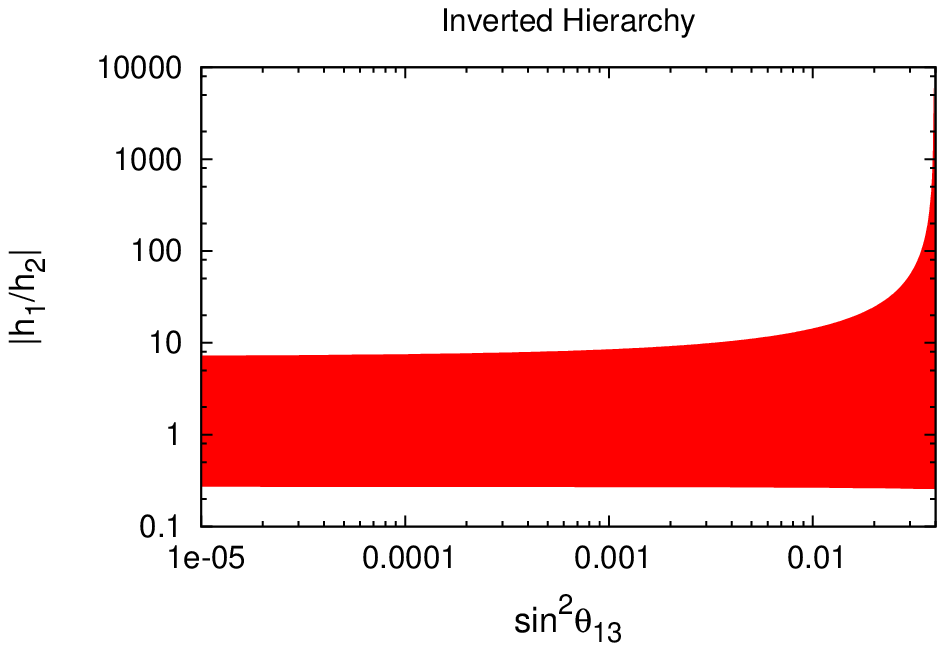}
	\caption{ The plot of the allowed value of $h_i/h_j$
	as a function of $\sin^2\theta_{13}$.}
	\label{fitting}
\end{figure}

The couplings $h_i$ will mediate $\ell_i \to \ell_j \gamma$ decays (see the next
section for detailed discussions). One has for the ratio of branching ratios,
\begin{eqnarray}
\frac{{\rm BR}~(\mu \to e \gamma) }{{\rm BR}~(\tau \to e \gamma)} = \left|
\frac{h_2}{h_3} \right|^2 \times {\rm BR}(\tau \to e\ol{\nu}_e \nu_\tau),
\label{br-ratio}
\end{eqnarray}
where ${\rm BR}(\tau \to e\ol{\nu}_e \nu_\tau)\simeq 0.18$ \cite{pdg}.
Eq. (\ref{br-ratio}) has an interesting consequence.
As explained above, in the NH case, $|h_2/h_3| \sim 1$. This means
the branching ratio of $\tau \to e \gamma$ cannot exceed $5.3 \times 10^{-11}$ because
of the limit on the branching ratio ${\rm Br}(\mu \to e \gamma) <2.4 \times
10^{-12}$ \cite{meg}. A measurement of ${\rm BR}~(\tau \to e \gamma)$ near the current
experimental limit of $\sim 10^{-8}$ would rule out the NH scenario. Of course, 
for these decays to have significant branching ratios, the leptoquarks must have
masses not much above a TeV.  In Fig. \ref{fitting}, we also show the ratio $|h_1/h_2|$
as a function of $\sin^2\theta_{13}$ allowed in the model for the NH case (lower left panel)
and IH case (lower right panel).  The ratio ${\rm Br}(\tau \rightarrow e\gamma)/{\rm Br}(\tau \rightarrow
\mu \gamma) = |h_1/h_2|^2$ in our model, which can server as a further test.

\section{Experimental constraints}

The new interactions shown in Eqs. (\ref{new_int}) and (\ref{new_int1})
can induce lepton flavor violation processes such as $\mu \to e \gamma$ and
$\mu \to 3e$ decays. In this section we analyze various such processes
and derive limits on model parameters. LHC experiments have set lower limits on
the leptoquark mass: $M_{1,2} > 376 \,(319)$ GeV for the first generation
leptoquarks and $M_{1,2} > 422 \,(362)$ GeV for the second generation leptoquarks,
assuming branching ratio of 1(0.5) \cite{ATLAS}. Our fit to neutrino mass suggests that
the branching ratio of the leptoquark to muons is about 0.5, so we shall adopt 
the corresponding limits in this section.  

\subsection{$\bm{\mu \to e \gamma}$}

This process occurs in the model via the one loop diagrams shown in Fig. \ref{mu-e-gamma}. There are
two couplings which are responsible for this process: $g_{ij}$ and $h_i$. In fact, the
predictions of this model are similar to the ones discussed in Ref. \cite{bj}, with one difference
that here we have interference between diagrams generated by $g_{ij}$ and those
induced by $h_i$.  In the present model, ignoring
the electron mass, which is an excellent approximation, the branching ratio is given by
\begin{equation}
{\rm BR} (\mu\ \to e \gamma) = \frac{27\alpha}{16\pi
G_F^2}\left|
F(x_{d_i})\frac{g^*_{1i}g_{2i}}{M_2^2} +
H(x_{u_\alpha})V^*_{\alpha 4} V_{\alpha 4} \frac{h^*_{1}h_{2}}{M_1^2}
 \right|^2,
\label{BR-mu-e-gamma}
\end{equation}
where $x_{d_i}\equiv m_{d_i}^2/M_2^2$ and $x_{u_\alpha}\equiv m_{u_\alpha}^2/M_1^2$.
The dimensionless functions $F(x)$ and $H(x)$ are given by \cite{hisano,lavoura}
\begin{eqnarray}
F(x) &=&
-\frac{x}{12}~\frac{(1-x)(5+x)+2(2~x+1)~{\rm
ln}~x}{(1-x)^4}, \nonumber \\
H(x) &=& -\frac{1}{12}~\frac{(1-x)(5~x+1)+2~x(2+x)~{\rm
ln}~x}{(1-x)^4}.
\end{eqnarray}
The branching ratios for other $\ell_i \to \ell_j \gamma$ processes can be derived in a similar way.
The resulting constraints on the model parameters are summarized in Table \ref{rare}. Here all 
of the experimental limits are taken from Ref. \cite{pdg} except for $\mu\to e \gamma$ limit
which is taken from Ref. \cite{meg}.

\begin{figure}[t]
\centering
	\includegraphics[scale=0.5]{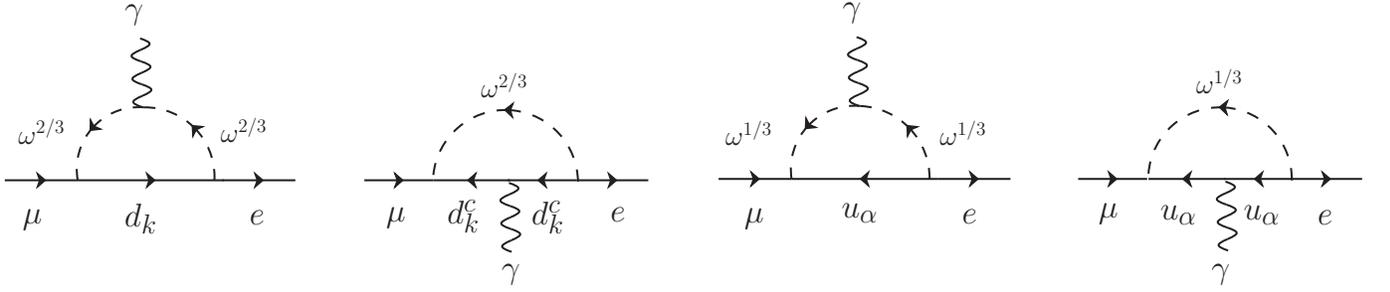}
	\caption{ One loop diagrams leading to $\mu \to e \gamma$ decay.}
\label{mu-e-gamma}
\end{figure}

An interesting feature of this analysis is that the $g_{ij}$
couplings are only weakly constrained from these processes.  This is
owing to a GIM--like cancelation for the amplitude for this process
from the first two diagrams of Fig. {\ref{mu-e-gamma}. This is similar to the model
discussed in Ref. \cite{bj}. This
cancelation occurs, in the limit of down quark mass being zero,
since the charge of the internal leptoquark
($2/3$) is twice as large and opposite in sign compared to the
charge of of the internal down quark ($-1/3)$.  The amplitude
for the diagram when the photon is emitted from the scalar line is twice smaller compared
to the diagram where it is emitted from the fermion line, which leads
to the cancelation. The amplitude that survives has a suppression of
$(m_b^2/M_{LQ}^2)$, which causes the weak limit.  Because of this cancelation, we can
derive correlated limits on the masses of the leptoquarks and the vector-like quark
from $\mu \rightarrow e\gamma$, since only the $h_i$ couplings are involved.
This is shown as a contour plot in Fig. \ref{mass}.  To get the largest possible
masses, we set the Yukawa couplings $h_1 = h_2 = 1$, as large as allowed by perturbativity.  If
$\mu \rightarrow e\gamma$ is discovered at the current limit \cite{meg}
the masses should lie to the left of the red contour
in Fig. \ref{mass}, while a measurement of ${\rm BR}(\mu \rightarrow e\gamma) = 1.0 \times 10^{-12}$
would require the masses to lie to the left of the blue contour.  The LHC reach for a 
leptoquark of this type is 1.5 TeV \cite{belyaev}, which would serve as a cross check in this case.

\begin{table}[h!]
\begin{center}
\begin{tabular}{cccc}\hline\hline
Process & BR limit & & Constraint \\ \hline \\ $\mu \to e \gamma$ &  $< 2.4
\times 10^{-12}$ & & $\left|
F(x_{d_i})\frac{g^*_{1i}g_{2i}}{M_2^2} +
H(x_{u_\alpha})V^*_{\alpha 4} V_{\alpha 4} \frac{h^*_{1}h_{2}}{M_1^2}
 \right|^2
< \frac{1.39\times 10^{-19}}{{\rm GeV^4}} $
\\\\
$\tau \to e \gamma$ & $< 3.3 \times 10^{-8}$ & &
$\left|
F(x_{d_i})\frac{g^*_{1i}g_{3i}}{M_2^2} +
H(x_{u_\alpha})V^*_{\alpha 4} V_{\alpha 4} \frac{h^*_{1}h_{3}}{M_1^2}
 \right|^2
< \frac{4.8\times 10^{-5}}{{\rm GeV^4}}$
\\\\
$\tau \to \mu \gamma$ & $< 4.4 \times 10^{-8}$ & &
$\left|
F(x_{d_i})\frac{g^*_{2i}g_{3i}}{M_2^2} +
H(x_{u_\alpha})V^*_{\alpha 4} V_{\alpha 4} \frac{h^*_{2}h_{3}}{M_1^2}
 \right|^2
< \frac{6.6\times 10^{-15}}{{\rm GeV^4}}$
\\\\
\hline \hline
\end{tabular}
\end{center}
\caption{{Constraints on model parameters from $\ell_i \to \ell_j
\gamma$.}} \label{rare}
\end{table}

\begin{figure}[h]
\centering
\includegraphics[scale=0.8]{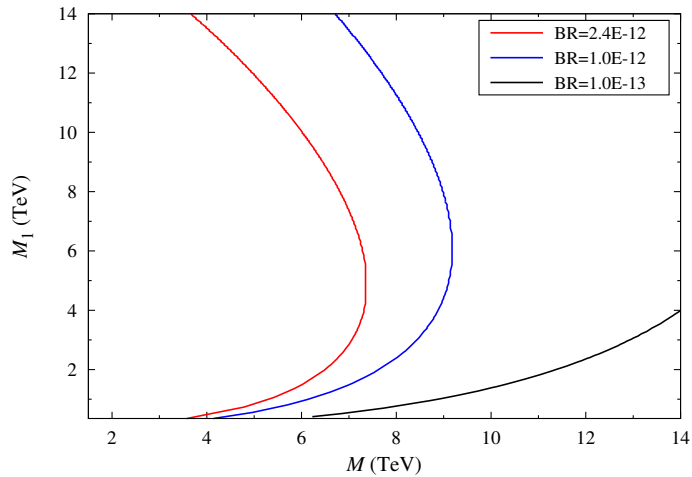}
\caption{The allowed region for for the leptoquark mass $M_1$ versus the vector-like quark mass $M$ from $\mu \to e \gamma$.
Here regions left of a contour is allowed for a fixed value of the branching ratio.  Thus, if $Br(\mu \rightarrow e\gamma) = 1.0 \times 10^{-12}$
is measured, the masses must lie to the left of the blue contour.}
\label{mass}
\end{figure}

\subsection{$\bm{\mu \to 3e}$}

In this process, the photon can be off-shell, and therefore, there is no GIM-like
cancelation for the $g_{ij}$ contributions. It turns out that in addition to the photon penguin diagrams,
there are also $Z$ penguin diagrams and box diagrams (the Higgs boson
exchange is suppressed by the electron mass). The mixing between vector-like
quark and SM chiral quarks also plays a role in this process. The expression for the decay width
is rather lengthy, which we do not present for brevity, but it is similar to the one given in Ref. \cite{bj}.   
To simplify the problem in deriving the constraints, we assume that only the top quark
mixes with the vector-like quark, or equivalently $s_{14},s_{24} \ll \lambda^3$ in Eq. (\ref{u-par}),
where $\lambda \simeq 0.22$ is the Wolfenstein parameter, while $s_{34}$ could be as large as
0.3, consistent with constraint from $Z\to b\ol{b}$ constraint \cite{alwall}.
For $\omega^{2/3}$ exchange
(corresponding to down-type quark inside the loop), we assume that there
is no accidental cancelation among the different couplings $g_{ij}$, and thus omit
terms such as $g_{13}g_{23}g_{jk}$ with $j,k=1,2$. For degenerate leptoquark masses of 400 GeV,
and for the vector-like quark mass set equal to 600 GeV, we obtain:
\begin{eqnarray}
|h_1h_2| < 2.7 \times 10^{-4} (3.4 \times 10^{-4}); \, \left| g_{13}g_{23}\right| < 1.7 \times 10^{-3};
\, \left| g_{1j}g_{2j}\right| < 8.6 \times 10^{-4},~j=1,2,
\end{eqnarray}
for $c_{34} = 0.98 \,(1.0)$.  These limits are obtained by assuming that contributions from one type
of coupling dominates at a time.  While these limits are stringent, they do not pose any restriction
on the neutrino masses and mixings.  The decay $\mu \rightarrow 3e$ may be within reach of next
generation experiments, with the couplings lying in the range $(10^{-2}-1)$ and the leptoquark mass
around a TeV.

\subsection{$\bm{\mu-e}$ conversion in nuclei}

Since this model features direct interactions of quark and
lepton via the leptoquarks, $\mu-e$ conversion in nuclei occurs. The diagrams are similar
to the ones discussed in Ref. \cite{bj}, with tree level and 
loop contributions.  There is a more direct link between neutrino mass and the loop
induce $\mu-e$ conversion process.   If we assume that only the top quark mixes with
the vector-like quark as in the case of $\mu\to 3e$, then there is no
tree level $\omega^{-1/3}$ exchange contribution to this process. Following the procedure
outlined in Ref. \cite{bj}, 
from the limit on $\mu-e$ conversion in $^{48}{\rm Ti}$, we obtain (for  $M_{LQ} = 400$ GeV, and a vector-like quark
mass of 600 GeV)
\begin{eqnarray}
|h_1 h_2| < 2.2 \times 10^{-4} (9.8 \times 10^{-3}); \,
\left| g_{13}g_{23} \right| < 8.7 \times 10^{-4}; \,
\left| g_{11}g_{21} \right| < 4.6 \times 10^{-6},
\end{eqnarray}
for $c_{34}=0.95\,(1.0)$.  Again, this analysis suggests that for natural values of the model parameters,
this process may be within reach of next generation experiments.

\begin{figure}
\centering
	\includegraphics[scale=0.55]{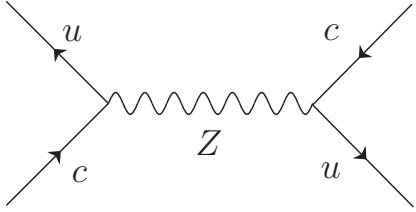}
	\caption{The tree level FCNC diagram leading to $D-\ol{D}$ mixing.}
\label{ddbar-mixing}
\end{figure}

\subsection{Tree level $\bm{D^0-\ol{D^0}}$ mixing}

The FCNC that occurs in the up--quark sector (see Eq. (\ref{nc-vec})) induces tree level $D^0-\ol{D^0}$ mixing 
mediated by the $Z$ boson, as shown in Fig. \ref{ddbar-mixing}. The neutral Higgs boson
induced contribution from Eq. (\ref{nc-sc}) is suppressed by light quark mass and can be ignored.
The mixing amplitude is given by \cite{branco95}
\begin{eqnarray}
M_{12}^{\rm new} = \frac{\sqrt{2}G_F}
{3}~\(V_{14}V^*_{24}\)^2
~m_D f_D^2 \hat{B}_D \eta_D(\mu),
\end{eqnarray}
where $m_D=1.9~{\rm GeV}$ is the $D$ meson mass, $f_D=0.201$ GeV is the $D$ meson decay
constant, $\hat{B}_D(\mu \sim m_D)=0.865$ is the bag parameter,
and $\eta_D(\mu \sim m_D)=0.78$ is the QCD correction factor. All the numbers here are taken
from Ref. \cite{bona08}. By using $\Delta m_D=1.6 \times 10^{-14}$ GeV \cite{pdg},
one obtains the constraint
\begin{equation}
\left| V_{14}V^*_{24} \right| < 2.5 \times 10^{-4}.
\label{D0}
\end{equation}
According to Eq. (\ref{u-par}) this constraint implies 
$|c_{14}s_{14}s_{24}| < 2.5 \times 10^{-4}$.  As a result of this limit, unlike in a four generation
model where there is no such FCNC process, the vector--quark contributions to meson mixing (eg., in the $B_d$ sector) cannot be too large.

\subsection{$\bm{B_s \to \mu^+ \mu^-}$ decay}

Recently, the CDF collaboration has reported a hint for the decay $B_s \to \mu^+\mu^-$, with
${\rm BR}=\(1.8^{+1.1}_{-0.9}\)\times 10^{-8}$ \cite{cdf-bsmumu}.  LHCb collaboration has not confirmed such a hint,
and quotes an upper limit $BR(B_s \rightarrow \mu^+\mu^-) < 1.4 \times 10^{-8}$ \cite{LHCb}.
The SM prediction for this branching ratio is ${\rm BR}(B_s \to \mu^+ \mu^-) = 3.2 \times 10^{-9}$,
which means there is ample room for new physics in this process.

Tree level exchange of leptoquarks can contribute to $B_s \rightarrow \mu^+ \mu^-$ decay in our model.  
The branching fraction is given by \cite{digheBs}
\begin{eqnarray}
{\rm BR}(B_s \to \mu^+\mu^-) = \frac{\left|g_{22}g_{23}\right|^2}{128\pi M_2^4}
\tau_{B_s}f_{B_s}^2 m_{B_s} m_{\mu}^2 \sqrt{1 - 4 \frac{m_\mu^2}{M_2^2}}
\end{eqnarray}
where the SM contributions have been ignored.  This enables us to fit ${\rm BR}(B_s \to \mu^+\mu^-) = 1.8 \times 10^{-8}$, 
with $M_2=400$ GeV for the leptoquark mass, and $\left|g_{22}g_{23}\right| \sim 4.2 \times 10^{-3}$.

\subsection{New CP violation in $\bm{B_{s}-\ol{B}_{s}}$ mixing}

Our model of leptoquarks and vector-like quark generates new contributions to $B_s-\ol{B}_s$ mixing.
There are two sources, one  through LQ induced box diagrams, and the other through SM-like
box diagrams with the vector-like quark (see Fig. \ref{bs-mixing}).  Including these contributions,
the $B_s-\ol{B}_s$ mixing amplitude becomes
\begin{eqnarray}
M_{12s} &=& \left\{
\frac{G_F^2m_W^2}{12\pi^2} \left[ \(V_{32}^*V_{33}\)^2 \eta_{33}S_0(x_3)
+ 2 \(V_{32}^*V_{33}\)\(V_{42}^*V_{43}\) \eta_{34}S_0(x_3,x_4)
+ \(V_{42}^*V_{43}\)^2 \eta_{44}S_0(x_4)
\right]
\right. \nonumber \\
&& \left. + ~ \frac{\left(g_{i2}g_{i3}^*\right)^2}{192\pi^2 M_2^2}\eta_B \right\}
~m_{B_s}f_{B_s}^2\hat{B}_s(\mu).
\label{dispersive}
\end{eqnarray}
The functions
$S_0(x_\alpha),S_0(x_\alpha,x_\beta)$ with $x_\alpha\equiv m_{u_\alpha}^2/m_W^2$
are the Inami-Lim functions \cite{inami-lim}, whereas $\eta_{ij},\eta_B$ with
$i,j=3,4$ are the QCD correction factors. The numerical values for these factors
for a 600 GeV for vector-like quark mass are \cite{soni-alok10}
\begin{equation}
\eta_{33} = \eta_B = 0.5765, \,\,
\eta_{34} = \eta_{44} = 0.514.
\end{equation}

\begin{figure}
\centering
	\includegraphics[scale=0.5]{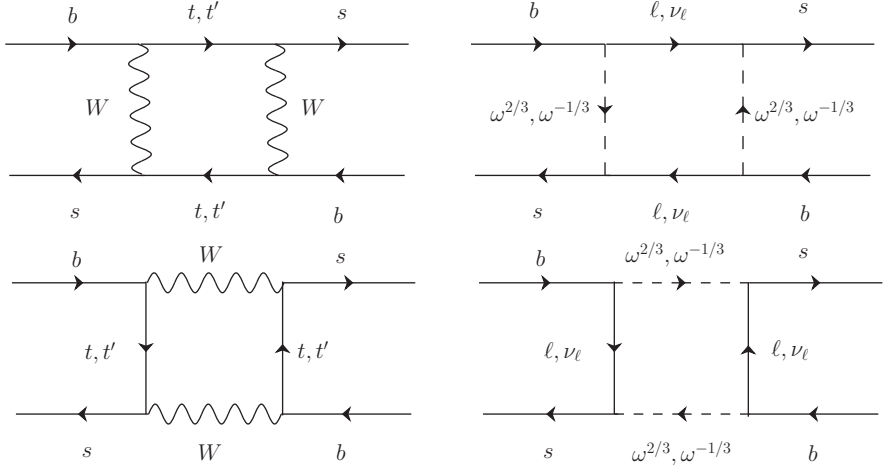}
	\caption{ Box diagrams leading to $B_s-\ol{B}_s$ mixing}.
	\label{bs-mixing}
\end{figure}

It is sometimes more convenient to parametrize $M_{12s}$ as \cite{ligeti10}
\begin{eqnarray}
\frac{M_{12s}-M_{12s}^{\rm SM}}{M_{12s}^{\rm SM}} &\equiv& r_{1s} e^{i2\sigma_{1s}} +
r_{2s} e^{i2\sigma_{2s}}
\label{dis-par}
\end{eqnarray}
where $\{r_{1s},\sigma_{1s}\}$ and $\{r_{2s},\sigma_{2s}\}$ are the new contributions.
With this parametrization, one can write
\begin{eqnarray}
\Delta m_{B_s} &=&  \Delta m_{B_s}^{\rm SM} \left| 1 + r_{1s} e^{i2\sigma_{1s}} +
r_{2s} e^{i2\sigma_{2s}} \right|,
\label{mass-diff}
\nonumber \\
S_{J/\psi\phi} &=& \sin\left[2\beta_s^{\rm SM} -
{\rm Arg} \(1 + r_{1s} e^{i2\sigma_{1s}} + r_{2s} e^{i2\sigma_{2s}}\) \right],
\label{beta-def}
\end{eqnarray}
with $\beta_s^{\rm SM} \equiv
{\rm Arg}\left[\(-V_{32}V_{33}^*\)/\(V_{22}V_{23}^*\)\right]= 0.019 \pm 0.001$,
$\Delta m_{B_s}^{\rm SM} = (19.3 \pm 6.74)~{\rm ps}^{-1}$.

The main reason to highlight this phenomenon is because there are
hints for new sources
of CP violation beyond the SM  in the D\O\ measurement inferred from the charge
asymmetry in the same sign di-muon decay of the $B$ mesons \cite{abazov}:
\begin{equation}
A^b_{sl} = \frac{N^{++}-N^{--}}{N^{++}+N^{--}} = -(0.787 \pm 0.172 \pm
0.093) \%. \label{mu-asym}
\end{equation}
Here $N^{++}(N^{--})$ is the numbers of events containing two $b$
hadrons that decay semileptonically into two positive (negative)
muons. Eq. (\ref{mu-asym}) can be written as a linear combination of
two asymmetries \cite{grossman07,abazov}
\begin{equation}
A^b_{sl} = (0.506 \pm 0.043) a^d_{sl} + (0.494 \pm 0.043) a^s_{sl},
\label{sl-asym}
\end{equation}
where $a^q_{sl}$ ($q\equiv d,s$) is defined as \cite{abazov}
\begin{equation}
a^q_{sl} = \frac{\Gamma(\ol{B}_q \to \mu^+X) - \Gamma(B_{q} \to
\mu^-X)}{\Gamma(\ol{B}_q \to \mu^+X) + \Gamma(B_{q} \to \mu^-X)}.
\label{wc-asym}
\end{equation}
In the SM, $a^d_{sl}=-4.8^{+1.0}_{-1.2} \times 10^{-4}$ and
$a^s_{sl}= (2.1 \pm 0.6)\times 10^{-5}$ \cite{lenz}, so that
$(A^b_{sl})^{\rm SM} = -2.3^{+0.5}_{-0.6} \times 10^{-4}$ which is
$3.9\sigma$ away from the current measurement (see Eq.
(\ref{mu-asym})).  A likely explanation is that there is a new source
of CP violation in $B_s-\overline{B}_s$ mixing.

Additionally, the measurements of relative phase between $B_s$ mixing
amplitude and $B_s \to J/\psi\phi$ decay amplitude
($S_{J/\psi\phi}\equiv \sin 2\beta_s^{J/\psi\phi}$)
as well as the measurements of decay width difference performed by
CDF \cite{aaltonen07} and D\O\ \cite{abazov08} yield \cite{hfag10}:
\begin{eqnarray}
\beta_s^{J/\psi\phi} &=& 0.47^{+0.13}_{-0.21} ~\cup~ 1.09^{+0.21}_{-0.13},
\label{beta-exp}
\end{eqnarray}
Here there is a $2.1\sigma$ discrepancy from SM prediction for
$\beta_s$, which may be another hint for physics beyond the SM.

It is interesting to see whether the vector--like charge $2/3$ quark can resolve these
problems. The best fit to the data for a fourth generation quarks, including the
preferred values of the CKM mixing angles, is given 
in Ref. \cite{dighe10} which shows $\beta_s = 0.03$ corresponding to $r_{1s}=0.02$
which is still far from the experimental central
value (see Eq. (\ref{beta-exp})).  This result should hold in the present model as well.
We conclude that the mixing with vector-like quark is not
enough to get the central value of $\beta_s$, so the LQ induced box diagram
is a more promising source for the new physics here.

Ignoring the effect of extra family mixing, the LQ contribution $\{r_{2s},\sigma_{2s}\}$,
can satisfy the best fit given in Ref. \cite{ligeti10}, i.e. $\{0.5,120^\circ\}$.
This would correspond to  $\left| g_{i2}g_{i3}\right| \sim 0.05$
where the index $i$ is summed. However, since the limits ${\rm BR}(B_s \to \mu^+\mu^-) \sim 10^{-8}$
and ${\rm BR}(B_s \to e^+e^-)< 2.8 \times 10^{-7}$ must be satisfied, the dominant contribution
should arise with the $\tau$ lepton inside the box diagram loop.  
The phase is not constrained and therefore can take the fitted value of
$120^\circ$.

\subsection{$\sin2\beta$  versus $\epsilon_K$ }

\begin{figure}[t]
\centering
	\includegraphics[scale=1.25]{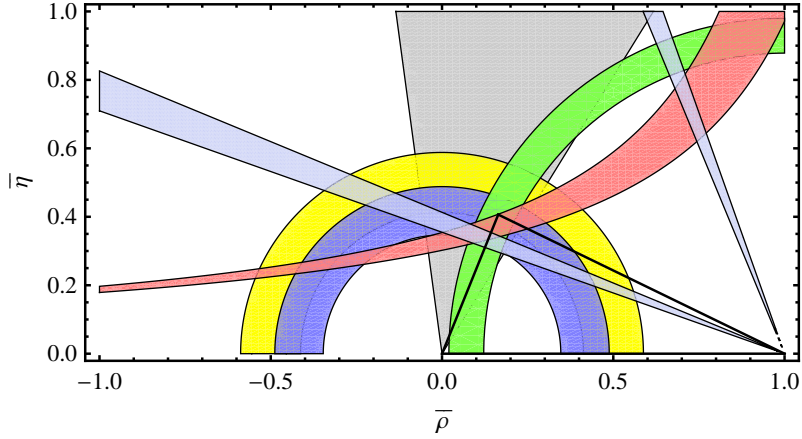}
	\caption{ The unitarity triangle fit leading to $\sin2\beta$
	determination.}
	\label{utfit}
\end{figure}

Analogous to the $B_s$ system, this model also provides new contributions to $B_d-\overline{B_d}$ mixing,
arising through LQ induced box diagrams and new mixing between SM quarks and the
vector-like quark.  Owing to the stringent constraint arising from the tree level $D^0-\overline{D^0}$ mixing, see
Eq. (\ref{D0}), analogous mixing in the $K^0$ system would be more suppressed.  The new contributions to the
$B_d$ mixing may explain the apparent discrepancy between $\sin 2\beta$ and $\epsilon_k$ determinations
within the SM.   The fitted value of $\sin 2\beta$
obtained from $|\epsilon_K|$, $\Delta m_d/{\Delta m_s}$, $|V_{ub}/V_{cb}|$,
${\rm BR}(B\to \tau\nu)$, and $\gamma$ measurements is given by
\begin{eqnarray}
\sin2\beta^{\rm fit} = 0.79 \pm 0.039
\end{eqnarray}
which differs by $3.1\sigma$ from the world average experimental value \cite{hfag10},
\begin{equation}
S_{J/\psi K_s} \equiv \sin2\beta^{\rm exp} = 0.671 \pm 0.023.
\end{equation}
The three family SM fit to the various CKM observables
is shown in Fig. \ref{utfit}.
A possible explanation is that there is new physics that affects the $B_d$ system, which we
parametrize as
\begin{eqnarray}
\frac{M_{12d}-M_{12d}^{\rm SM}}{M_{12d}^{\rm SM}} &\equiv& r_{1d} e^{i2\sigma_{1d}}
+ r_{2d}e^{i2\sigma_{2d}},
\label{bd-dis-par}
\end{eqnarray}
where
\begin{small}
\begin{eqnarray}
M_{12d} &=& \left\{
\frac{G_F^2m_W^2}{12\pi^2} \left[ \(V_{31}^*V_{33}\)^2 \eta_{33}S_0(x_3)
+ 2 \(V_{31}^*V_{33}\)\(V_{41}^*V_{43}\) \eta_{34}S_0(x_3,x_4)
+ \(V_{41}^*V_{43}\)^2 \eta_{44}S_0(x_4)
\right]
\right. \nonumber \\
&& \left. + ~ \frac{\left(g_{i1}g_{i3}^*\right)^2}{192\pi^2 M_2^2} \eta_B \right\}
~m_{B_d}f_{B_d}^2\hat{B}_d(\mu).
\label{bd-dis}
\end{eqnarray}
\end{small}
This is analogous to the discussion of $B_s$ mixing.  
With this formula, one can write
\begin{eqnarray}
S_{J/\psi K_s} &=& \sin \left[ 2\beta^{\rm fit} + \phi^{B_d} \right],
\end{eqnarray}
where
\begin{eqnarray}
\phi^{B_d} = {\rm Arg} \left[ 1 + r_{1d} e^{i2\sigma_{1d}} + r_{2d} e^{i2\sigma_{2d}}
\right].
\end{eqnarray}

Unlike in the $B_s$ system, the effect of the vector-like quark can induce
a significant effect to resolve the tension in $\sin2\beta$ determination. In order
to see this effect, let us ignore for the moment the LQ contributions. Then, if we choose
\cite{dighe10}
\begin{eqnarray}
V_{31}^*V_{33} &=& 0.009 e^{i0.56} \nonumber \\
V_{41}^*V_{43} &=& 0.00096 e^{-i1.35},
\end{eqnarray}
we obtain $S_{J/\psi K_s} = 0.68$, which is in the good agreement with the experimental value. 
Note, that the LQ contribution is not strongly constrained
by neutrino mass nor lepton flavor violation, so it can get close to the experimental value,
as long as the LQ mass is less than 1 TeV.

\subsection{Neutrinoless double beta decay}

Although this model can accommodate inverted hierarchy in which
neutrinoless double beta decay ($\beta\beta_{0\nu}$)  may occur with sizable
effect, it is still interesting to see that even in the normal hierarchy case,
such process may be observed
through vector-scalar exchange mechanism \cite{babu-mohapatra}, depicted in
Fig. \ref{beta-decay}.

\begin{figure}
\centering
	\includegraphics[scale=0.6]{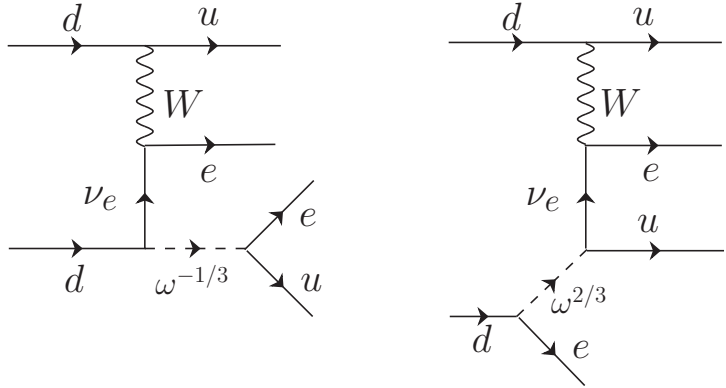}
	\caption{ Diagrams leading to the neutrinoless double beta decay
	through vector-scalar exchange.}
	\label{beta-decay}
\end{figure}

The effective Lagrangian of the new $\nu-e^c-u-d$ vertex after Fierz
rearrangement is
\begin{equation}
\mathcal{L}_{\rm eff}^{\rm new} =
\frac{G_F}{4\sqrt{2}}~\epsilon~\left[\ol{u}\(1+\gamma_5\)d~
\ol{\nu}_e\(1+\gamma_5\)e^c ~ + ~
\frac{1}{2}\ol{u}\sigma^{\mu\nu}\(1+\gamma_5\)d ~
\ol{\nu}_e \sigma_{\mu\nu}\(1+\gamma_5\) e^c \right]
\end{equation}
where
\begin{equation}
\epsilon = \frac{g_{11}^*h_iV_{14}^*}{2\sqrt{2}M_1^2G_F}
\left(1-\frac{M_1^2}{M_2^2}\right).
\end{equation}
This process is similar to MSSM models without $R$-parity violation discussed in
\cite{babu-mohapatra} and \cite{hirsch99}. Following Ref. \cite{hirsch99},
and by using the results from Heidelberg-Moscow experiment on $\beta\beta_{0\nu}$
decay rate \cite{heidelberg-moscow}, one obtains for $M_1=300$ GeV and $M_2=350$
GeV
\begin{equation}
\left| g_{11}^* h_1 V_{14}^* \right| \leq 4.3 \times 10^{-7}.
\end{equation}
The mixing matrix elements $|V_{14}V_{24}|$
is constrained by $D-\ol{D}$ mixing process and has to be less than $10^{-4}$.
Since the coupling $g_{11}$ is not constrained by neutrino mass, it could be of order one.
For $V_{14} \lesssim 10^{-5}$ and $h_1 \sim 10^{-2}$ (assuming NH case and vector mass of order sub-TeV) from lepton flavor violation constraints,
one ses that neutrinoless
double beta decay might be observable even in the case of normal mass hierarchy.  Of course, for this
to be valid, the leptoquarks and vector-like quark have to be light.

\section{Conclusions}

In this paper we have presented a new two-loop neutrino mass generation model
which has the effective operator ${\cal O}_3$ of Eq. (\ref{op}).  Generating
this effective operator in a renormalizable theory would require the addition
of a charge 2/3 vector-like quark and a scalar leptoquark doublet tot the
standard model spectrum.  We have studied the phenomenology of this model.
This model can explain the CP violation parameters in the $B_s$ and the
$B_d$ system.  The leptoquarks of the model generate new CP violating contributions
in $B_s-\overline{B}_s$ mixing, which can explain the di-muon anomaly reported
by D\O\.  The apparent tension in the determination of $\sin2\beta$ from $B_d$
decays and from the global analysis including $\epsilon_K$ from the $K$ meson
system also finds a natural explanation in this model.  
Neutrinoless double beta decay may occur through vector-scalar exchange and may
be observable even with a normal hierarchy in the neutrino masses.

\section*{Acknowledgement}
This work is supported by in part the US Department of Energy, Grant Numbers DE-FG02-04ER41306.

\end{document}